**Accepted Version**

Publication date: January 2023

Embargo: no embargo

European Union, Horizon 2020, Grant Agreement number: 857470 — NOMATEN — H2020-WIDESPREAD-2018-2020

DOI: https://doi.org/10.1016/j.jallcom.2022.168196

# Microstructure and mechanical properties of mechanically-alloyed CoCrFeNi high-entropy alloys using low ball-to-powder ratio

A. Olejarz [a,*], W. Y. Huo [a,b,*], M. Zieliński [a], R. Diduszko [a,c], E. Wyszkowska [a], A. Kosińska [a], D. Kalita [a], I. Jóźwik [a], M. Chmielewski [a,c], F. Fang [d], Ł. Kurpaska [a]

[a] - NOMATEN Centre of Excellence, National Centre for Nuclear Research, A. Sołtana 7, 05-400 Otwock-Świerk, Poland

[b] - College of Mechanical and Electrical Engineering, Nanjing Forestry University, Nanjing, 210037, China

[c] - Lukasiewicz Research Network, Institute of Microelectronics and Photonics, 32/46 Al. Lotników Str, Warsaw, 02-668, Poland

[d] - Jiangsu Key Laboratory of Advanced Metallic Materials, Southeast University, Nanjing, 211189, China

* Corresponding authors: E-mail address: artur.olejarz@ncbj.gov.pl (A. Olejarz), wenyi.huo@ncbj.gov.pl (W. Y. Huo)

**Highlights**

- With a low ball-to-powder ratio, CoCrFeNi HEAs were produced using mechanical alloying followed by spark plasma sintering.

- Elevated mechanical properties were obtained by adjusting the milling time.

- Increased milling time resulted in structural homogeneity and improvement of mechanical properties.

**Abstract**

High-entropy alloys are extensively studied due to their very promising properties. However manufacturing methods currently used to prepare HEAs are complicated, costly, and likely non-industrially scalable processes. This limits their evolution and poses questions regarding the material's applicability in the future. Considering the abovementioned point, we developed a novel methodology for efficient HEA production using a low ball-to-powder ratio (BPR). Using different milling times, we manufactured four HEA powder precursors using a BPR of 5:1, which were later sintered via the Spark Plasma Sintering technique and heat treated. Microstructural characterization was performed by optical microscopy, Scanning Electron Microscopy equipped with EDS and EBSD detectors, and X-ray diffraction. Mechanical properties were measured using nano and microhardness techniques. In this work, we follow the structural evolution of the material and connect it with the strengthening effect as a function of milling time. Furthermore, we discuss the impact of different sintering and annealing conditions, proving that HEAs characterized by high mechanical properties may be manufactured using low BPR.

1. Introduction

Over the last two decades, the popularity of high entropy alloys (HEAs) in scientific environments has significantly increased, mainly due to confirmation of their excellent properties over a wide range of temperatures [1–4]. The most heavily studied cases involve chemical composition changes or the addition of other non-metallic elements [4–6]. The goal of these investigations addresses one particular problem: the promotion of the strengthening mechanisms that are responsible for high strength, especially at high temperatures. Many different chemical compositions and structures have already been evaluated by conducting a series of complex simulations and experiments. For example, some works have revealed that small composition changes cause a significant increase in yield strength [7–9] and corrosion resistance [9,10] or may enhance the radiation resistance of the system [11,12]. Despite a number of available works, the choice of the best manufacturing technique remains ambiguous.

One of the HEAs' most popular manufacturing methods is arc melting [7,9]. However, this technique allows the synthesis of materials on a small scale, which may be problematic for industrial manufacturing. As an alternative to the arc melting technique, the classical powder metallurgy route should be considered promising due to its higher availability and cost-effectiveness [1,12]. One must remember that powder metallurgy is a relatively broad term. Therefore, two subsequent processes composing this technology must be distinguished. The first step leads to the preparation of the powder mixture. The most frequently used is mechanical alloying, where powder particles are mixed in a particular vessel followed by milling in, for example, a planetary ball milling device. During this process, the powders undergo welding and fracturing [1]. The described procedure results in, at first, obtaining a finely divided powder with a modified structure and random distribution of elements [13,14]. Proper material formation occurs during compaction as a second process step.

Among many different methods of powder consolidation, the spark plasma sintering (SPS) technique deserves special attention [15–17]. In this process, the internal heating source, which always occurs during compression of powder particles, is enhanced by external heating introduced through the graphite die by direct current. Therefore, target materials can be formed significantly below the melting temperature of each composing element. The most significant advantage of this method is a more prominent shape-forming perspective. It has also been proven that SPS is shorter and much cheaper than other synthesis techniques [18]. These three reasons may be vital in filling many technological gaps. For this reason, the primary efforts should be turned into understanding the process parameters to obtain a material with expected elevated functional properties.

Many different parameters should be regarded during mechanical alloying to obtain an effective milling process. Among them, the ball-to-powder ratio (BPR) takes a special place. BPR is the ball-to-powder-weight quotient. Many works related to HEAs manufacture produce them with a high BPR – equal to or over 10:1 [19–23]. Relatively rarely, which is surprising, is the scientific literature discussing lower BPRs. One must admit that materials produced with higher BPR usually present promising properties, although the volume of the obtained powder remains the main problem in this approach [24]. With high BPR, less powder can be milled in one campaign. Consequently, the amount of the powder may be insufficient for synthesising large items, or many milling campaigns need to be performed. Thus, inappropriate powder chemical compositions may be created. This significantly increases the chances of forming the non-uniform structure after sintering. This may lead to deterioration of the mechanical properties of the materials and quick failure of the sintered item (for example, through cracking along poorly sintered or weaker strength part of the material). This is considered the most significant challenge in implementing this alloy manufacturing method at an industrial scale. On the other hand, it is essential to remember that a higher BPR value results in higher milling process efficiency, reducing the milling time. This leads to the conclusion that different trade-offs between BPR and milling time have to be tested for each alloy composition or at least the alloy family.

More often, the sintering parameters or the heat treatment procedure attract more attention. For example, Manta et al. [25] and Raja Rao with Sinha [26] analysed the impact of sintering temperature on the microstructure and mechanical properties of HEAs and oxide dispersion strengthened (ODS) HEAs. They found that parameters such as yield strength or hardness reach their maximum values at approximately 1000 °C and then significantly decrease. Moreover, different phases may form depending only on the sintering temperature [25]. The post-sintering heat treatment process is also often used for sample homogenisation. Moravcik et al. [27] demonstrated that annealing results in a hardness decrease due to the dissolution of the σ-phase in the metal matrix. In addition to standard questions related to sintering time and temperature, an interesting but not intuitive factor is the process control agent (PCA). PCA is a fluid used during milling to restrain cold welding between powder particles. Jayasree et al. [28] analysed the milling process under three conditions: toluene, methanol, both with the same initial C-content, and dry milling. Recorded XRD patterns from powder milled with toluene present a more prominent $Cr_7C_3$ peak than in the methanol-supported sample. These examples show that the modification of the manufacturing method and parameters may considerably impact properties despite the same chemical composition of the final material.

Taking into account all these considerations, this paper aims to analyse the impact of milling time – the most crucial manufacturing factor due to the possibility of a wide range of control. The milling time can be easily extended or reduced, resulting in ineffective or very efficient powder alloy preparation. Milling pauses prevent cold welding and provide a better mixing effect. In our work, the BPR was limited to 5:1 to restrain the impact of this parameter. As mentioned above, BPR reduction causes lower efficiency but elevated productivity. Consequently, this may lead to finding a more effective manufacturing path when other process parameters, such as milling time, are set correctly. The more alloyed powder produced in one milling cycle significantly expands the perspectives for manufacturing more advanced shapes and larger elements.

The milling time of the HEA powders was analysed many times, but the description of the impact on mechanical parameters of the sintered samples is missing. For this reason, in our work, we analysed different manufacturing parameters with respect to the properties of the final materials. In this study, an optical microscope, SEM coupled with EDS and EBSD detectors, and X-ray diffraction techniques were employed to analyse the structural properties of manufactured materials. Microhardness tests and nanoindentation were used to assess the mechanical properties of materials on two different scales. This work aims to comprehensively describe the effects occurring during milling. The second goal of our work is to evaluate their impact on sintered and annealed samples. This analysis is the first step toward providing a detailed description of how to modify manufacturing process parameters if one wants to obtain materials with satisfactory mechanical properties.

## 2. Experimental

### 2.1. Sample preparation

In this work Co (<1.6 μm, purity: 99.8 %), Cr (<45 μm, purity: 99.0 %), Fe (<10 μm, purity 99.9 %), and Ni (<37 μm, purity: 99.8 %) produced by Alfa Aesar were mixed in equiatomic composition under an Ar atmosphere and set up in a WC jar with Φ5 mm WC balls. Powders were subsequently milled in the Retsch PM100 high-energy ball milling system. A milling speed of 250 rpm was used during the entire powder milling process. The mixing process was performed in 15 min intervals to prevent powder consolidation. In addition, n-heptane was added to avoid cold welding, as suggested by Liu and Cheng [20,29]. The cumulative milling process was carried out at different times: 10, 20, 30, and 40 h. The ball-to-powder ratio was always 5:1. After powder milling, the spark plasma sintering (SPS) technique was employed for powder consolidation. The sintering temperature was 950 °C. Specimens were heated at the rate of 100 °C/min. The holding time was 8 min at the target temperature, and sintering was performed by applying 50 MPa of pressure. The dimensions of the samples after sintering were 10 mm in diameter and height. After the sintering process, the samples were mechanically ground to remove the graphite layer. Heat treatment was performed in a muffle furnace at 1050 °C for 12 h. The heating rate used in the process was 100 °C/h, and the chamber was flushed twice with high purity argon. After heat treatment, all samples were water quenched.

After the sintering and annealing process, the samples were ground with sandpapers: #320 to #2500, followed by polishing with diamond suspensions of 9 μm, 3 μm, and 1 μm. Polishing was performed until a mirror-like effect was reached, and then electropolishing using a mixture of 8% perchloric acid and 92% ethanol was performed. The electropolishing process was conducted under 25 V for 30 s at room temperature. The final step of preparation before SEM observations was ion polishing. This was done by using the PECS II (Gatan) device. For clarity, notation of the synthesised and homogenised samples has been proposed and is given in Tab. 1.

Tab. 1 Specification of the investigated specimens.

| x | 10 h | 20 h | 30 h | 40 h |
|---|------|------|------|------|
| **After sintering** | S-10 | S-20 | S-30 | S-40 |
| **After annealing** | A-10 | A-20 | A-30 | A-40 |

## 2.2. Density and microstructure observation

Density measurements were carried out using the Archimedes method on a RADWAG X2 Series. Measurement was repeated at least five times for each sample. An Olympus BX53M optical microscope - equipped with a 100x lens was used to illustrate the microstructure changes in the sintered samples. Furthermore, a ThermoFisher Helios 5 UX Scanning Electron Microscope was employed for detailed structural characterisation. EDS analysis was carried out to reveal the elemental composition of the phases developed during the sintering and after annealing processes. EBSD maps presenting the evolution of the grain size distribution and the grain orientation with increasing milling time in annealed samples were obtained. To estimate the average grain size from EBSD maps, OIM Analysis 8 by EDAX software was used.

### 2.3. X-ray diffraction

Structural studies were carried out on a Bruker D8 Advance diffractometer using *Cu K$_α$* radiation (λ = 1.54056 Å) and the Bragg-Brentano geometry. The diffraction patterns were collected in the 2θ range from 40° to 100° at room temperature on a LYNXEYE XE-T detector working in high-energy-resolution 1D mode with an opening angle of 2.941°. The Bruker DIFFRAC.TOPAS program, which is based on fundamental parameters profile fitting (FPPF) and Rietveld approaches, was used to refine the models of the identified phases (via JCPDS-ICDD PDF-2 2021 [30]) to describe the experimental diffraction patterns.

### 2.4. Hardness

The hardness of specimens was measured using a hardness tester with a mounted Vickers type indenter by applying a 100 g load and 12 s dwell time. At least six indents on each sample were made. Nanoindentation was performed using a NanoTest Vantage system from Micro Materials. A Synton-MDP diamond Berkovich-shaped indenter was used in this study. Tests were conducted using a 50 mN load. Thirty-six indentations were performed on each sample with 30 μm spacing between the indents. The Fisher-Cripps relation was used to calculate the relationship between Vickers microhardness $H_V$ and Berkovich nanohardness $H_B$, which was done by employing eq. (1) [31]:

$$H_V \left[\frac{kg}{mm^2}\right] = 94.495 H_B \quad (1)$$

### 3. Results

### 3.1. Density and microstructure observation

The densities of the samples measured before and after annealing are presented in Tab. 2. Recorded data show no milling time dependence. In the case of sintered samples, the densities of the specimens exceed 95% of the material's theoretical density. This suggests that the consolidation process was efficient, and we obtained satisfying densification levels. The theoretical density of the material was calculated based on equation (2) [32]:

$$\rho = \frac{4 m_{aAvg}}{V_{cell}} \quad (2)$$

, where $m_{aAvg}$ corresponds to the averaged atomic mass of Co, Cr, Fe, and Ni, and $V_{cell}$ is the volume of an FCC unit cell calculated with the lattice constants derived from XRD data. In this case, the average weight volume of the FCC unit cells is considered. The theoretical density was estimated as 8,23 g/cm$^3$.

Tab. 2 also shows the density loss of the annealed samples in reference to the sintered samples. The observed decreases are insignificant and the results still exceed 90% of the theoretical density calculated from sintered samples. The recorded density loss is most likely related to the annealing under the active atmosphere. The effect is less significant for the A-30 and A-40 samples.

Tab. 2 Densities of the specimens sintered from powders mixed for 10, 20, 30, and 40 h before and after annealing.

| Density [g/cm$^3$] | 10 h | 20 h | 30 h | 40 h |
| --- | --- | --- | --- | --- |
| After sintering (S) | 7.96 | 7.86 | 7.89 | 7.84 |
| After annealing (A) | 7.33 | 7.36 | 7.56 | 7.54 |

The LM and SEM images of the sintered samples are presented in Fig. 1. Recorded images show visible porosity. This is particularly detectable for the S-10 and S-20 specimens presented in Fig 1 (a) and (b), respectively. The porosity can be observed as darker points in LM images – Fig. 1 (a) and (b) - marked with white circles. It disappears or is significantly reduced (beyond the detection level) for the samples characterized by longer milling times – S-30 and S-40 (Fig. 1 (c) and (d), respectively). In addition, different areas similar to oval-shaped particles are visible in Fig. 1 (a-d) and (e-h) and under magnification in Fig. 2 (a). In general, the number of these particles per unit area decreases with increasing milling time.

The SEM-EDS technique was employed to evaluate the elemental composition of the matrix and the oval-shaped particles before annealing. At least five points from the matrix and five points from the particle of each sample were analysed. In the matrix phase of the S-40 sample, all four elements are distributed almost uniformly. A small Cr depletion may be observed. The chemical composition analysis of particles revealed high Cr saturation with traces of other elements. Recorded data show that the Cr content reaches 85 wt%. Detailed results are collected in Tab. 3, suggesting Cr-rich phase formation in the sintered samples. The obtained EDS maps confirm this elemental distribution. This is presented in Fig 2 (b). EDS point analysis of the matrix (as marked in Fig. 2 (a)) confirmed the increasing homogeneity as a function of milling time, which is presented in Fig. 2 (c). The results from optical microscopy and SEM-EDS suggest the presence of an undesirable chromium enriched phase. The crystallinity of this phase may depend on the starting chemical composition, but the process parameters may also impact it.

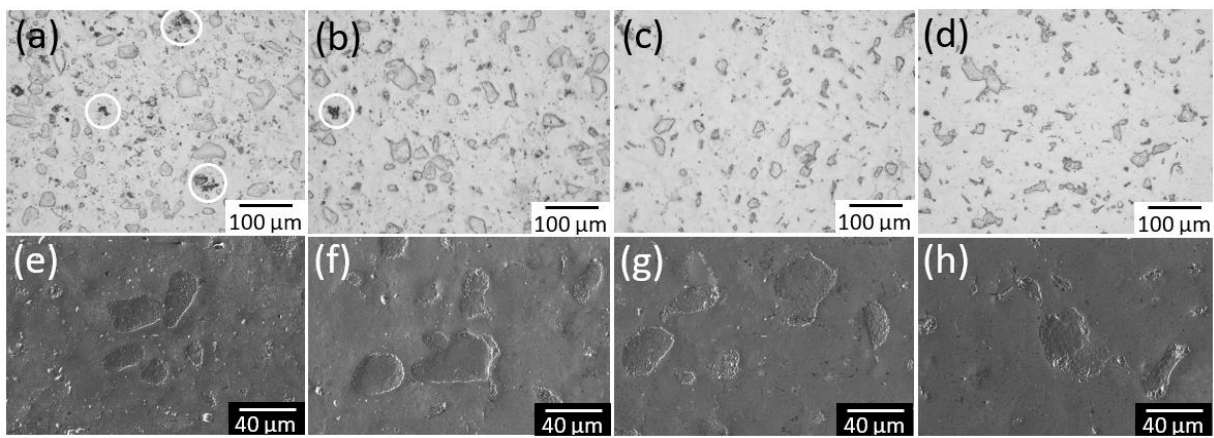

Fig. 1 Microstructure of manufactured HEA with different milling times S-10 – S-40 h from OM (a-d) and SEM (e-h). Marked with white circles areas in Fig. 1(a) and (b) represent detected pores.

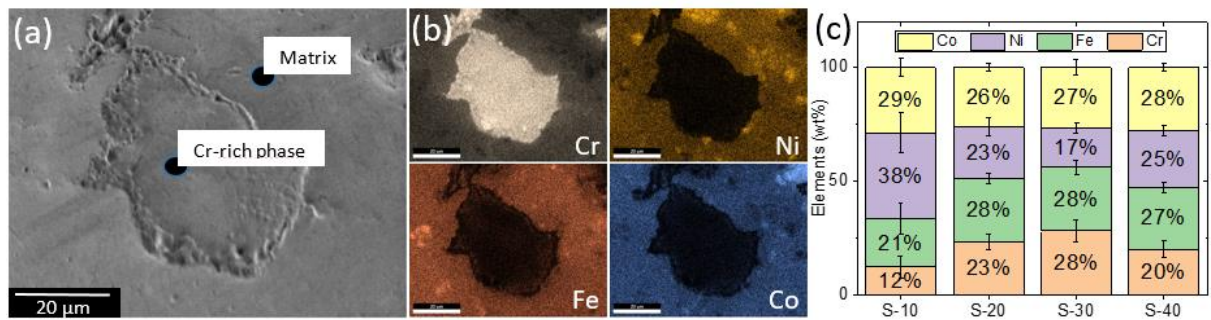

Fig. 2 (a) SEM image of Cr-rich phase after sintering with marked EDS detections spots (b) EDS maps, and (c) quantitative (wt%) EDS results recorded during point analysis of the matrix.

Tab. 3 Chemical composition of the matrix and Cr-rich phases after sintering (S-40 sample).

| Element | Co | Cr | Fe | Ni |
|---|---|---|---|---|
| Matrix [wt%] | 27.8 | 20.2 | 27.2 | 24.8 |
| Cr-rich phase [wt%] | 3.5 | 90.7 | 3.7 | 2.1 |

After annealing, the samples show a more uniform structure without pores detected after sintering. To better understand and visualise the microstructure of the studied samples after annealing, SEM/EBSD analysis was performed. In backscattered electron SEM image (Fig. 3 (a)), small particles have been revealed. These particles have a different shape than those after sintering. They are jerkier, and their size is generally smaller than those after sintering. EDS point analysis and EDS mapping were again employed to understand the chemical composition of the particles and the surrounding matrix. The investigation confirmed the presence of a Cr-rich phase. The obtained results are collected in Tab. 4. Recorded EDS maps demonstrate a highly uniform distribution of all four elements in the matrix. Only slight chromium depletion can be noticed. Extending the milling time results in a more uniform chromium distribution in the matrix. This phenomenon is presented in Fig. 3 (b). At the same time, the chromium-rich particles show small contents of other elements. Generally, particles are partially dissolved in the matrix during heat treatment. One can observe that only the largest particles remain. Some small pores surrounding Cr-rich particles can also be noticed.

Tab. 4. Chemical composition of the matrix phase and Cr-rich phase after annealing (A-40 sample).

| Element | Co | Cr | Fe | Ni |
|---|---|---|---|---|
| Matrix [wt%] | 25.50 | 24.98 | 25.76 | 23.76 |
| Particle [wt%] | 5.13 | 82.47 | 8.45 | 3.95 |

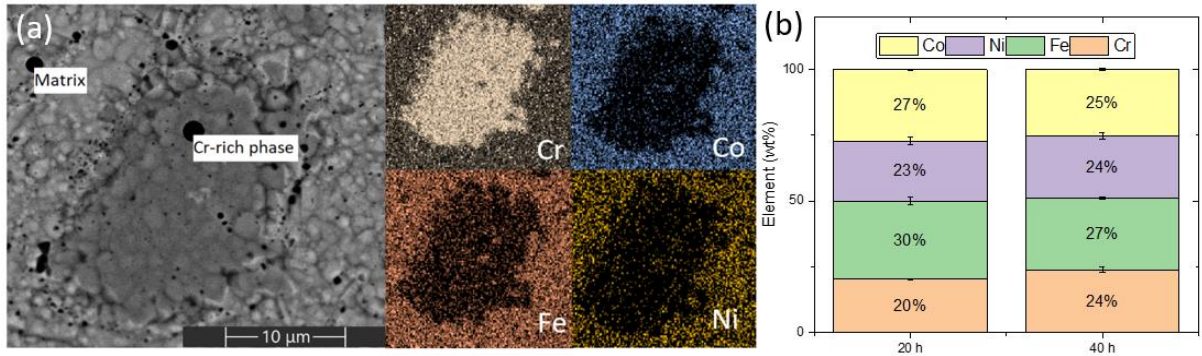

Fig. 3 (a) SEM image of the representative Cr-rich phase obtained after annealing at 1050 °C for 12h (left-hand-side) and recorded EDS maps of the Cr-rich phase and its nearest neighbourhood matrix, (b) wt% elemental composition of matrix as a function of time measured for specimens milled during 20 and 40h.

The grain size as a function of milling time is presented in Fig. 4. One can observe that a fine-grained structure dominates in annealed specimens. The obtained maps demonstrate that the grain size decreases in the annealed samples as the milling time of the powders increases. The most considerable grain size reduction is visible between the A-10 and A-20 samples. A further increase in the milling time does not significantly impact the grain size after annealing. Applying the shortest milling time (10 h), a bimodal grain size distribution may be observed after annealing with larger grains surrounding smaller agglomerates (marked on the A-10 image with black circles and arrows). One can see that samples submitted to longer milling times present a more uniform grain size. Typically, for a polycrystalline material, a random orientation of the grains, with a high level of grain boundary misorientation, is observed.

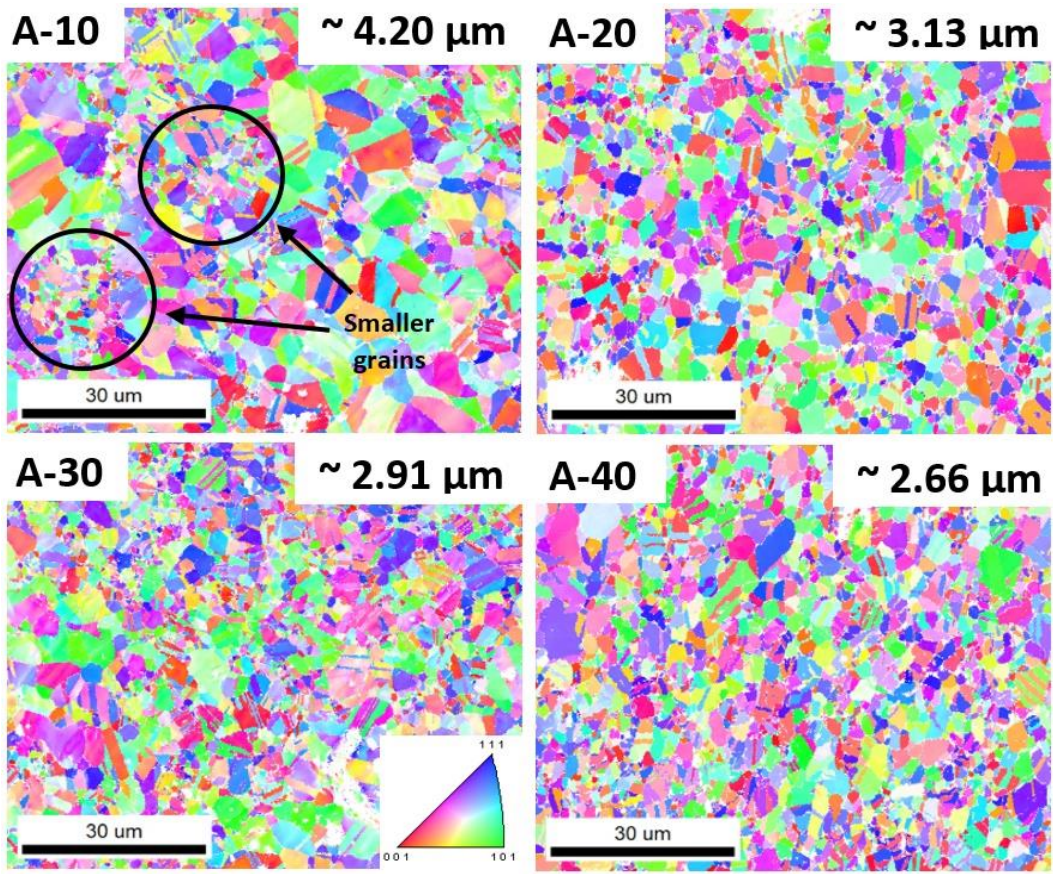

Fig. 4 EBSD orientation maps (IPF Z) of the annealed samples. Specimens were sintered and annealed from powders prepared with different milling times, from 10 to 40h.

### 3.2. X-ray diffraction

To evaluate the structural evolution during milling, XRD analysis was carried out on the samples before sintering. As expected, phases representing each metal used for alloy manufacture were identified. Moreover, our studies revealed the presence of the FCC + HCP cobalt structure. The volume-weighted average crystallite size decreases which makes the peaks broader and makes the impression of gradual peaks disappearance. However, even after the longest milling time, peaks can still be distinguished. The diffractograms of milled powders are presented in Fig. 5.

Fig. 6 (a) presents the XRD patterns recorded on the samples after sintering. Peaks representing input elements almost disappear, and new FCC phases are formed (marked with black triangles). One can see that every sample consists of a few FCC structures. This is revealed in Fig. 6 (b, c), where the convolutions of individual FCC peaks in S-10 and S-40 are described. A comparison of the results of the analysis of the X-ray diffractograms is presented in Fig. 7 (a). The total content of the FCC-type phases increases with increasing milling time. Notably, with the increasing milling time, the major FCC structure is promoted at the expense of the others, and their content gradually fades away. In Fig. 7 (b), the lattice parameters of different FCC phases as a function of milling time are compared. The three FCC phases have similar lattice parameters. In the S-40 sample, only two of them might be separated. The parameter of the fourth phase clearly deviates from the others. In shorter milled samples it increases, but it significantly decreases in S-40. The lattice parameter of this phase is close to that of Ni. FCC4 gradually dissolved as the milling time was extended.

Indeed in all of the specimens multiple, but similar, kinds of crystal domains exhibiting FCC ordering were observed, and they constituted the largest part of each specimen. The respective diffraction peak profiles can be modelled in a satisfactory manner only by incorporating several FCC-type contributions characterised by similar values of the crystal lattice constant "a" but varying volume-weighted mean crystallite size values. Such a situation is typical for specimens of alloys in which there is a lack of composition and/or size homogeneity of crystal domains (forming then grains). Still the composition and the size of each crystallite belong to a finite, although broad, range. Therefore, it is justified to claim that the reported FCC structure models are just representatives of the population indeed present in the specimens, and they serve well as contributions to the simulated scattering model. In consequence, the crystal lattice constant "a" of the FCC-type crystal domains actually present in the specimens accommodates all values within the reported limits rather than only the reported discrete values. The composition of each single crystallite is indeed different, although it may be regarded as "similar" within a chosen acceptance threshold. Moreover, there is a broad range of crystallites sizes in each specimen, which is also reflected in the EBSD images of the annealed samples.

As a consequence of the shapes of the diffraction peaks corresponding to FCC structures that required the choice of multiple FCC crystallite models, it was impossible to reliably assess both the volume-weighted mean crystallite sizes and any strain or stress of the crystal lattices at the same time.

To overcome this issue, the whole diffraction peak broadening was assumed to originate from the crystallite size effect. Although it has not been denied that the crystal structure of any phase can be defective, no estimation of crystal lattice strains or stress was provided, following the fact that investigation of defects was not the purpose of this research.

Clearly, the mechanical alloying process did not allow to obtain homogenous powders. However, the structures of the crystal phases found in the milled powders prove the progressive mixing of the elements with the extension of the milling time. As further evidence, the sintered materials exhibit improved homogeneity, i.e., one type of structure dominating over a lower number of other reported structure types, correlated with the increasing milling time of the precursors to the SPS process.

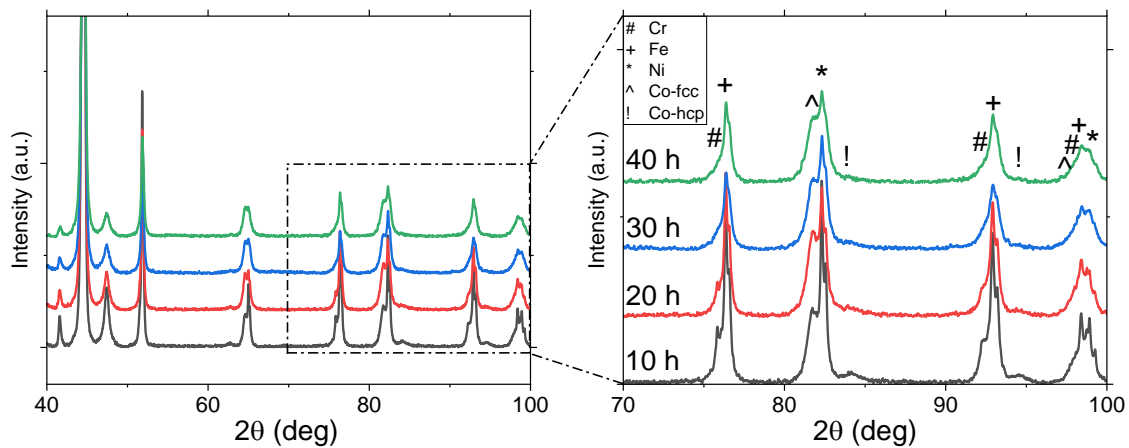

Fig. 5. (a) XRD patterns of powders milled for 10, 20, 30, and 40 h recorded in the range from 40 to 100 2θ (deg), (b) magnification of the range 70 – 100 2θ (deg) with marked characteristics for each element peak and phase.

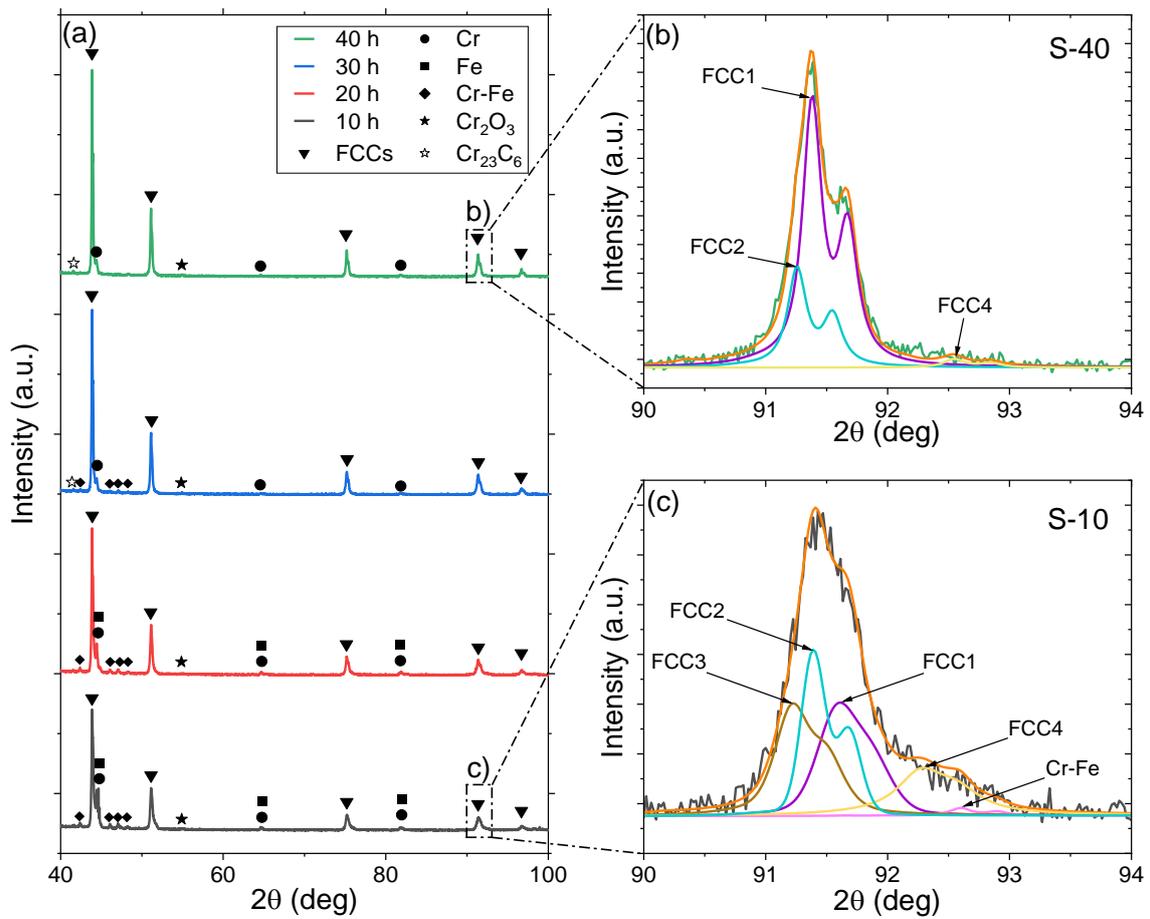

Fig. 6 (a) XRD patterns of sintered samples prepared from powders milled for 10 to 40 h; (b) (311) convolution of S-40; (c) (311) convolution of S-10. Both (b) and (c) graphs showing multiple FCC contributions.

In addition to the FCC phases, other structures were identified in the sintered samples. XRD analysis revealed trace amounts of input elements, intermetallic Cr-Fe phase, chromium oxides, and carbides. Carbides and oxides should be closely related to process limitations and cannot be avoided. The presence of other metallic/intermetallic phases suggests low efficiency of the milling process. The percentage of pure chromium and the Cr-Fe phase gradually decreases with increasing milling time, as presented in Tab. 5.

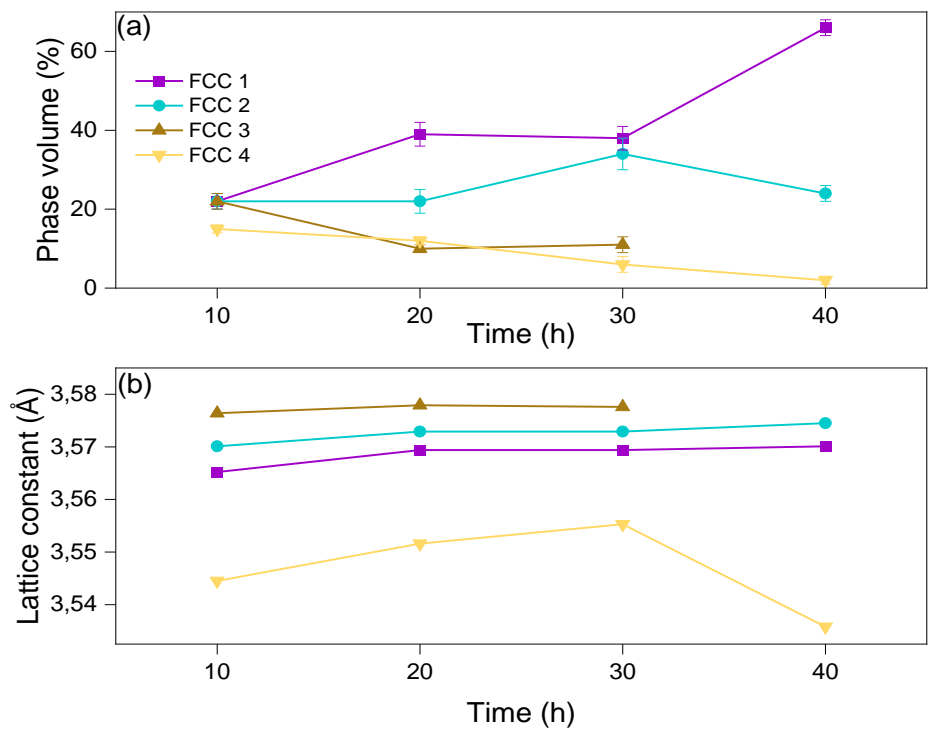

Fig. 7 (a) FCC volumes with error bars and (b) lattice parameters in the sintered samples as a function of the powder milling time (measurements uncertainties do not exceed 0.0012 Å for FCC3 and 0.0005 Å for others).

Tab. 5 Remaining phases in the prepared samples.

| x | S-10 | S-20 | S-30 | S-40 |
|---|---|---|---|---|
| Cr [%] | 8 | 8 | 3 | 3 |
| Cr-Fe [%] | 8 | 6 | 3 | 0 |
| Fe [%] | 1 | 1 | 0 | 0 |

The XRD patterns of the annealed samples are presented in Fig. 8. As a result of the heat treatment, diffraction peaks in the characteristic positions for the metal FCC phases may be observed, together with hardly visible peaks originating from a small amount of $Cr_2O_3$. A small content of chromium oxide is related to the presence of an active atmosphere during annealing (marked with a 'black star' in Fig. 8). The peaks of the Cr-rich phase identified from the microstructure analysis (see Fig. 3 (a)) were not detected due to its low content. Detailed analysis of the FCC-related peaks revealed deviations from the profiles corresponding to a perfect FCC structure with $a$ = 3.576 Å, which would be the best single-crystalline-phase fit to the experimental data. The first deviation is a slight and varying shift of the peaks positions towards smaller 2θ angles. The second is the asymmetry expressed by a left-hand shoulder, indicating an extra scattering contribution. Hence, each FCC-related peak has to be fitted with two sets of analytical functions corresponding to two crystalline domain types. We expect that the basic FCC lattice present in each specimen accumulates defects in some regions of the specimen volume causing residual strains of the crystal structure and, thus, a locally "deformed FCC" structure. Such fine structure results are similar to those obtained using synchrotron X-ray diffraction and neutron diffraction [33]. It implies that the HEA may not be a perfect solid solution [34], although EBSD results show that it has a single-phase structure.

Effectively, the X-ray scattering occurs as if there is a two-phase structure of the specimens, even if these phases are not physically separated by a grain boundary. The lattice parameters of the FCC phase at approximately 3.576 Å, regardless of the milling time, are consistent with the lattice parameters recorded after sintering. Additional scattering contribution that is believed to originate from a slightly distorted FCC phase.

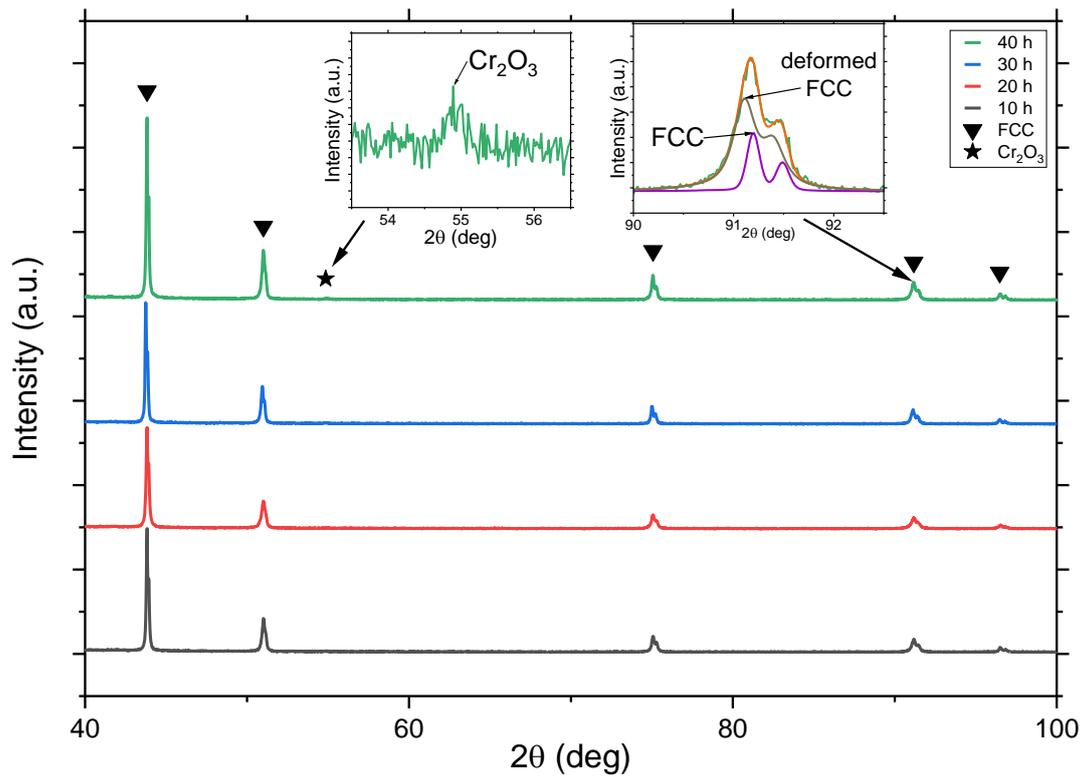

Fig. 8 XRD patterns of annealed samples prepared from powder milled for 10 to 40 h. Magnifications of the diffraction signal near 2θ = 55° and 91° indicate the presence of $Cr_2O_3$ and contributions to the peak profile.

### 3.3. Mechanical parameters

In Tab. 6, the microhardness of the sintered and annealed samples is presented. One can see that an increase in the milling time results in a hardness increase. This effect can be observed for both microhardness and nanohardness tests, suggesting the uniform distribution of the mechanical properties at both scales. The average microhardness of S-40 reaches 337 $HV_{0.1}$, whereas, for S-10, only 268 $HV_{0.1}$ was measured. Samples after heat treatment show lower hardness values. However, an increase in this parameter with milling time is clearly observable. It is worth noting that the sample submitted to the longest milling time shows the lowest hardness reduction after annealing of all samples and maintains it above 280 $HV_{0.1}$. At the same time, a significant increase in sample homogeneity may be observed.

Fig 9 (a) presents the average nanoindentation curves recorded on the annealed samples. One can observe the displacement of the L-D curves towards higher depths with decreasing milling time. All indentations were made with a maximal load of 50 mN. This resulted in the deformation of approx. 650 to 850 nm of the material. Given that the plastic deformation developed under the indenter tip (depending on the material) is considered to be even 10x higher than the maximal indentation depth, one can conclude that in each case, the recorded signal represents the response of at least a few grains. The previous paragraph (Fig. 4) shows that the average grain size varies from 4.2 to 2.6 µm. Performing the nanoindentation test allowed us to evaluate the heterogeneity of the mechanical properties in the sample volume. Due to the presence of a Cr-rich phase, the average nanohardness was estimated only for a matrix phase.

Mesoscale interfaces in materials, such as grain boundaries, play a significant role in determining the mechanical properties of materials. The recorded hardness increase can be correlated with the grain size decrease. This is consistent with the Hell-Patch effect [35,36], which provides a robust but empirical relationship between the macroscopic yield strength and average grain size in a material. Thus, the reported results suggest that we should observe an increase in yield strength with decreasing grain size. These tests are planned for the future to confirm this statement.

A comparison between the micro- and nanohardness values is displayed in Fig. 9 (b). The trends of both curves are similar. The hardness measured by the nanoindentation technique is slightly higher than the results from microhardness tests. The differences between them oscillate at approximately 15-20 %. This points to the conclusion that hardness at both scales is not size-dependent. Defects related to the surface roughness may impact the nanohardness data. Additionally, grain boundary misorientation, boundary energy, or different slip transfer rules between the grains may affect recorded values. Finally, one can see that local small Cr-rich precipitates may be detected regardless of the milling time and post-milling processes. These precipitates will block plastic deformation (dislocation movement), thus elevating the recorded hardness value.

Tab. 6 Microhardness test results

| Hardness [HV 0.1] | 10 h | 20 h | 30 h | 40 h |
| --- | --- | --- | --- | --- |
| After sintering (S) | 268 ± 42 | 321 ± 35 | 319 ± 16 | 337 ± 27 |
| After annealing (A) | 218 ± 63 | 250 ± 53 | 254 ± 28 | 282 ± 23 |

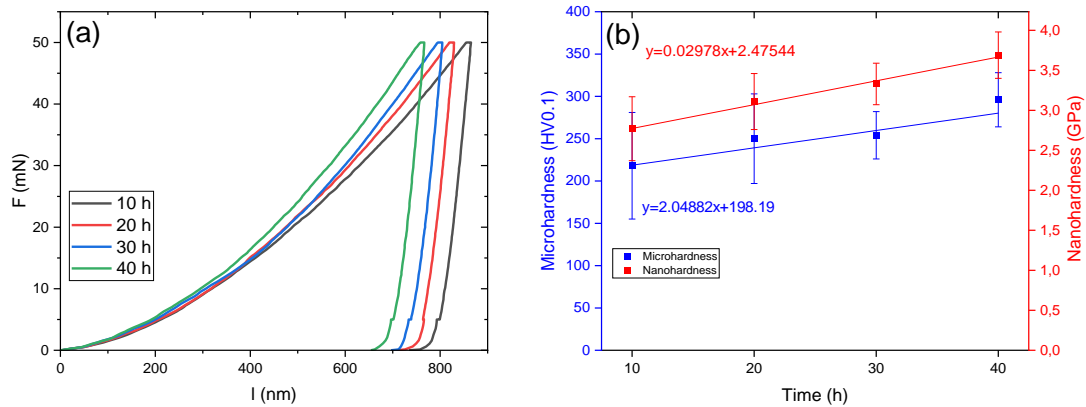

Fig. 9 (a) L-D curves recorded for all samples during the indentation test; (b) Micro- and nanohardness data comparison measured on the annealed samples.

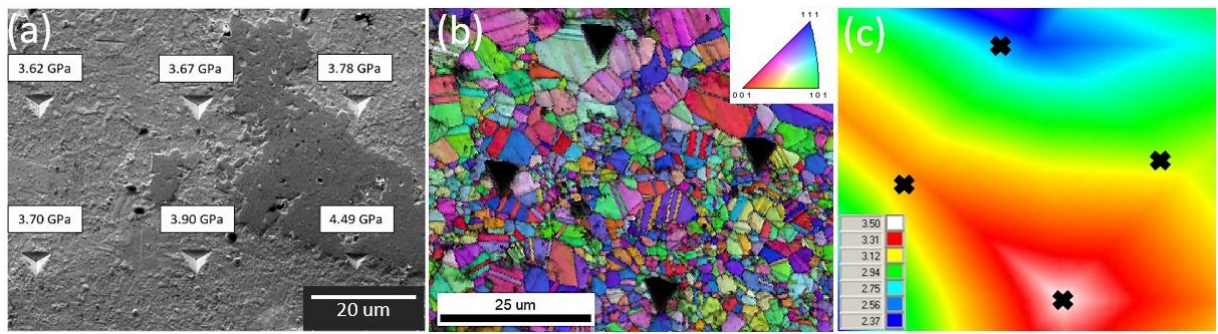

Fig. 10 (a) Nanohardness distribution near the Cr-rich particle; (b) EBSD orientation map demonstrating regions with different grain sizes; (c) average nanohardness distribution map from (b)

A visible increase in the standard deviation of microhardness after the heat treatment was recorded. This was observed especially on shorter milled samples. Individual indents from nanoindentation tests were visualised and analysed to better understand this effect. For this purpose, indents in the vicinity of Cr-rich particles are presented in Fig. 10 (a). The recorded SEM image with marked nanohardness reveals higher nanohardness of particles in comparison with the matrix phase. When moving away from the particle, the hardness of the matrix decreases. The impact of grain size is also worth considering. Especially in the shorter-milled samples, the framing of the smaller grains groups by the larger ones was evident. To understand this, the regions of smaller and larger grains were considered. The EBSD map with marked indents and nanohardness distribution are presented in Fig. 10 (b) and (c), respectively. The analysis shows the bimodal grain structure. The areas with smaller grain sizes present significantly higher hardness when compared with the larger ones. The nanohardness distribution map reflects the grain size distribution well. Even on a small area, the hardness difference may reach 1 GPa (~100 HV). These two effects explain (at least partially) discrepancies in the nanohardness results after the heat-treatment process.

## 4. Discussion

This paper examines the evolution of the microstructure and mechanical properties of HEAs produced by MA and sintered with the SPS technique. Consolidated powders were prepared with different milling times. In addition, this work focuses on HEA production by using a lower than that commonly seen in the literature BPR (5:1). The densities of the samples are collected in Tab. 1. The recorded results indicate no significant impact of the milling time on the material density. A high densification level was obtained in all samples. However, a slight density reduction was recorded after the heat-treatment process. Microstructure observations from sintered samples presented in Figs. 1 and 2 have revealed that Cr-rich particles are evenly distributed in the sample volume. Notably, the number of Cr-rich particles per unit area decreases with increasing milling time. At the same time, the porosity gradually disappears with milling time. After 30 hours of milling, the microstructure of the sintered samples seems to be without pores. The annealing process promoted the gradual dissolution of Cr particles. Only the largest particles remained, which can be seen in Fig. 3. EBSD orientation maps from Fig. 4 present grain refinement as a function of milling time, while the grain sizes are more uniform. XRD patterns of milled powders confirm the presence of primary metals used for consolidation in the form of pure phases. With increasing milling time, the respective diffraction peaks become broader, which is visible in Fig. 5. At presented in Figs. 6 and 7, the X-ray diffraction data show multiple FCC structures with a few chromium phases, chromium oxides, and carbides. Some intermetallic Cr-Fe phases formed after sintering have also been detected. As a function of milling time, the FCCs content increased at the expense of the Cr-Fe and Cr phases. Notably, the main FCC phase can be detected in the samples with a longer milling time. After heat treatment, the X-ray diffraction patterns (Fig. 8) present significant homogenisation, which was observed in all samples. Left-hand shoulder and shifts of FCC peaks may be observed. The micro-and nanohardness comparison in Fig. 9 shows slight differences between the results. However, the mechanical properties measured at both scales present increasing trends with increasing milling time.

**4.1. Structural transformation in mechanically alloyed and sintered samples**

Increasing the milling time increases the probability of ball-powder contact. As the number of contacts increases, more powder particles begin to deform or fracture [14,30]. Initially, all elements can be easily distinguished by XRD. As Kumar et al. [15] reported, the fracturing-welding trade-off is more balanced during longer milling times. Consequently, primary phases gradually disappear. HCP-Co disappears first due to the brittleness of the HCP structure [37]. In the case of powder-milled for the longest time, the effect of HCP-Co is almost invisible. During milling, Fe is gradually dissolved, followed by Cr. This is due to the differences in their melting temperatures and atomic bonds

After sintering, a few FCC phases are formed. The three (or two in S-40) structures have lattice constants similar to those of CoCrFeNi HEAs (0.357 nm), as detected previously by XRD in [6,19,38]. This effect is visible regardless of the manufacturing method. First (FCC1) may originate from the FCC-Co structure as revealed by XRD powder analysis. The assumption that the main phase is built on the FCC-Co structure results from a similar lattice parameter of the element to the phase obtained after sintering. The second could result from the BCC->FCC Fe phase transformation at 912 °C. The next FCC may originate from different FCC-Co with different lattice parameters obtained due to HCP-> FCC transformation at 430 °C [39]. After 40 hours of milling, the FCC3 dissolved. This corresponds well with the gradual disappearance of the HCP peak during milling (see Fig. 5 (b)). After 40 h, the HCP-Co-like phase dissolves and diffuses to other FCC phases, especially to the main FCC-Co structure. The content of the main FCC phase gradually increases after FCC3 disappears. The last phase may form on the FCC-Ni structure with a lower lattice constant than FCC-Co. The Ni powder used in a process has a considerably larger average size, which hinders atomic diffusion. Other larger elements deform the lattice. In the beginning, the lattice parameter slightly increases, but after 30 hours, it distinctly decreases. This indicates the migration of other elements to the matrix. The lattice constant of FCC4 in the 40-hour milled sample is similar to that of pure Ni. As a function of milling time, the main FCC is promoted. Atoms move from different structures to the main FCC1. After 40 hours, the main phase states approximately 2/3 of the sample composition. EDS analysis confirmed increasing homogeneity with increasing milling time.

The presence of almost-pure chromium suggests low efficiency of the milling process. Our results point to the conclusion that the chromium particles have remained in a matrix due to the highest melting point [15] and the strongest atomic bonds. The most significant chromium particle size among all powders used may also significantly impact chromium residuals in the microstructure. In our opinion, the energy delivered during milling was too low to break out particles to allow atomic diffusion. Recorded data suggest that the Cr-Fe phase was also formed. Due to the same space group, mutual dissolution is simplified [37]. Recorded data indicate that prolonged milling time forces the diffusion of elements to the main FCC phase, which is presented by the gradual disappearance of both phases.

The structural analysis revealed chromium carbide $Cr_{23}C_6$ and chromium oxide $Cr_2O_3$. They result from the manufacturing process. The appearance of chromium carbides is related to the usage of n-heptane ($C_7H_{16}$) as a process control agent and/or consolidation in a graphite die. Carbides form during the sintering process, and due to diffusion at high temperature, carbon atoms bond with chromium due to their high affinity [40]. This is directly related to the lowest number of valence electrons. The Gibbs free energy necessary to build the compound with chromium is the lowest. The appearance of this phase in longer milled samples must be linked with an increased diffusion coefficient of chromium during milling time. Oxygen appears in the material, especially after the sintering process due to surface oxidization [41]. Oxygen reacts with chromium significantly faster than with other elements. Chromium atoms migrate from the bulk towards the surface and form chromium oxides. The formation of these oxides on the surface may slightly disturb the chromium content, as detected by EDS/XRD techniques.

The analysis of annealed samples revealed the significant homogenisation of produced samples. The Cr-rich particles were still visible by SEM/EDS observation, but the number of particles was too low to be detected by XRD analysis. The difference between the lattice parameters of FCC phases known from sintered samples completely disappeared, and only a single FCC phase was detected on each sample. However, shifts towards lower 2θ and left-handed shoulders of FCC peaks effects may be observed. As no other diffraction peaks were observed, and only the additional effects close to the FCC positions can be seen (which make the experimental peaks asymmetric), it is a strong suggestion supporting the hypothesis that a distortion of the FCC phase has occurred. One must keep in mind that the elements used in this work have similar but not equal atomic diameters, and the fact that there is still a small presence of Cr-rich particles, near which chromium diffusion is easier than at longer distances confirms our hypothesis. Both effects may have an impact on not negligible peaks distortions. More research to elucidate this phenomenon is planned in the future.

Li et. al. [17] observed similar crystal phases using higher BPR (10:1). Comparing with the results presented in our work, the significant differences are (i) The Cr-depleted matrix phase, with chromium oxides and carbides precipitates, was denoted with a wide range of grain sizes, but the average grain size was smaller and (ii) there was lack of big particles of the Cr-rich phase [17,42,43]. These two facts show that afract higher BPR makes the process more efficient. It can be concluded that higher BPR determines faster atom diffusion to the main matrix phase, the grain size refinement ultimately resulting in higher homogenisation in a shorter period.

## 4.2. Strengthening improvement

The mechanical properties of the manufactured samples were compared with literature data. Different results representing materials obtained by casting methods [44–46] and SPS [47] are given in Tab. 7. Despite a comprehensive literature check-up, only a few studies have reported the nanohardness of alloys with this chemical composition. This is surprising, as this alloy is considered a starting point to design advanced HEAs. The A-40 sample shows similar hardness to arc melted specimens prepared by Huo et al. [46]. Although a concise milling time was used, the material developed by our group presented encouraging results in reference to cast samples. One should remember that the higher nanohardness in the samples reported in this paper could also originate from grain refinement forced by the sintering process.

The material produced by Laurent-Brocq et al. [47] has a bimodal structure with two different grain sizes. Our study confirms this finding, however, in our case, large grains were detected after 10 h of milling while small grains were detected after 40 h. To better understand the effect of the sintering process, the Cantor alloy synthesized by MA+SPS was considered. The presented results again indicate the higher hardness in the four element materials despite an insufficient milling process. In this case, introducing Mn at the expense of other elements, especially Cr and Fe, may result in a slight nanohardness reduction.

Tab. 7 Nanohardness comparison from different literature data.

| Chemical composition | Condition | Nanohardness | Source |
|---|---|---|---|
| CoCrFeNi A-40 | MA+SPS | 3.69 ± 0.29 GPa | This work |
| CoCrFeMnNi | Vacuum Induction Casting | 2.69 ± 0.06 GPa | [44] |
| CoCrFeNi | Arc Melting | 2.71 ± 0.13 GPa | [45] |
| CoCrFeNi | Arc Melting after tensile test | 3.40 – 3.80 GPa | [46] |
| CoCrFeMnNi | MA+SPS | 3.20 ± 0.20 GPa | [47] |

Higher Cr dissolution causes an increase in sample homogeneity. Fine powder promotes changes in the strengthening mechanism. With increasing milling time, the strengthening mechanism evolves from grain size + particle strengthening to grain size + solid solution strengthening in sintered samples due to the gradual distribution of Cr, Fe, and Ni atoms in a lattice of the FCC-Co matrix phase. The heat treatment forces the homogenisation of the structure by dissolving the residual phases and promoting the dominant FCC. Solid solution strengthening leads to lattice distortion due to atomic size differences. Lattice distortion is one of the most expected effects in HEAs. A higher standard deviation in shorter milled samples suggests a lack of homogeneity. The presence of larger and smaller grains may impact the variation in the microhardness results recorded on the annealed samples. In addition, the Cr-rich phase presents a much higher hardness than the matrix, which affects the apparent hardness increase near Cr-rich particles.

An evident difference from the presented results is the heterogeneity of the samples manifested by a much higher standard deviation on annealed samples. A typical feature of arc melted materials is a high homogeneity with relatively low mechanical properties, as presented in Tab. 7. In the present work, we show that the Cr-rich particles partially dissolved in the matrix during annealing. At the same time, the chromium atoms diffused from particles into the matrix. The highest saturation of the lattice with the chromium atoms occurred at a short distance from the Cr-rich phase. As particles move away, the chromium content decreases, reflecting a slightly lower hardness. It is worth noting that in longer milled samples (A-30 and A-40), the discrepancies in the nanohardness results are greatly related to the gradual chromium dissolution already performed during milling. At the same time, areas with smaller and larger grains may be observed. This is most likely due to uneven grain growth during the heat treatment process. The strength of a sample increases with grain size reduction. The example of a bimodal structure presented above (see Fig. 9 (b-c)) perfectly reflects this relationship. Higher saturation with chromium and other atoms in the matrix and greater grain refinement are the main factors determining the increase in mechanical properties and improved homogeneity. The presented results suggest that both parameters may be improved with milling process improvement.

In the case of annealed samples, micro/nanohardness increments of approximately 33% from 10 to 40 hours (see fig. 9b) are observed. On the assumption that the yield strength and hardness are proportional [47,48], the Hall-Petch rule was employed to define the value of the grain boundary strengthening effect on the total strengthening phenomenon. The theoretical value of strengthening improvement was estimated to be 26% covering almost all the micro/nanohardness improvement. Both values indicate that the hardness increase in thermally-treated samples is mainly due to grain size refinement. Many reasons may cause the remaining discrepancy, i.e., other strengthening effects, such as twin boundaries, strengthening by oxides detected by XRD, surface roughness and even measurement errors.

Our observations are consistent with Li et al. [17], who demonstrated that for the same chemical composition and very close manufacturing process with a higher BPR, the main factor determining the strength of the material is grain boundary strengthening. The detected microhardness was slightly higher than that of the A-40 sample. The higher microhardness coincides with the lower average grain size, which is a consequence of a higher BPR (10:1). The trend of grain size evolution as a function of milling time in the present paper suggests that a better strengthening effect may be achieved if other process parameters are optimized.

**4.3. Strategy for synthesis using low BPR**

Improving the manufacturing process is a significant challenge to obtain materials with better properties. BPR reduction may be an exciting way to obtain more powder for further synthesis, which may be particularly important for industrial applications. This work demonstrates that reducing BPR results in a drop in milling process efficiency in some cases. In the literature, a few examples of manufacturing a single FCC phase HEA by powder metallurgy can be found [29]. A common phenomenon is the presence of at least two different phases and artifacts in the form of carbides and oxides [49–51]. However, in those cases, a higher BPR was used. Our data suggest that in some instances, BPR reduction should result in milling time elongation to receive good material. The main evidence for an insufficient milling process with low BPR is the appearance of large unreacted Cr particles. A more fragmented fraction facilitates chromium diffusion to the matrix phase, hindering Cr-rich particle and intermetallic phase formation. More extended milling promotes one FCC phase at the expense of the other phases. However, even the longest time used in this work was still inadequate to ensure an efficient process. On the other hand, process elongation causes lower productivity, and some other unexpected phases may be formed. Spark plasma sintering is a rapid process that does not allow chromium diffusion. Consequently, mistakes made during milling cannot be overcome by sintering. However, producing other high-entropy alloys with lower BPR may still be possible when elements with higher diffusion rates are considered. Finding the milling time/BPR trade-off is vital in obtaining the optimal milling process parameters. Another solution worth checking is increasing the rotation speed, which should induce more powder-ball and powder-powder contacts.

The difference in powder particle size in mechanical alloying may also cause some problems. The Co and Fe powders were more fragmented at the beginning of the process, while the maximum diameters of Ni and Cr were 37 μm and 45 μm, respectively. Fine powder facilitates atomic diffusion and, therefore, higher quality of the sintered sample. Additionally, increasing the total ball mass might determine more effective fragmentation and diffusion. Another solution may be pre-milling – milling of all larger powder fractions to obtain similar granulation before the main mechanical alloying starts. Suprianto et al. [52] applied a similar approach, and the results were very promising. However, the economic factor must be considered in such a case. One must remember that double milling process employment causes significant process elongation. It is also possible to control the size of the pure elemental powders at the moment of their production when the HEA manufacturer can order elemental powders with specific characteristic from the suppliers.

Moreover, the presented results confirmed the significant homogenisation of the samples during the heat treatment process. Long annealing at high temperature forced diffusion and homogenisation. However, the main disadvantage of this solution is a hardness loss (<20 %), which is related to grain growth at high temperature.

5. **Conclusions**

In summary, the impact of milling time and low BPR on the sintered and annealed CoCrFeNi high-entropy alloy samples were analysed. This paper aims to connect the milling process by using a reduced BPR with the material's microstructural features and mechanical properties.

1. The presented results confirm that the milling time improves the homogeneity of the material by promoting a more uniform distribution of the elements in the sintered samples.
2. Dissolution of the Cr-rich phase in the matrix has been observed.
3. The heat treatment used allowed sample homogenisation. These, in consequence, promote a bigger FCC lattice distortion effect.
4. Milling time elongation induces grain size refinement. According to the Hall-Petch rule, the recorded hardness increase is close to theoretical strengthening improvement.
5. Similar dependence in samples milled with higher BPR has been reported previously.
6. The biggest obstacle in producing low BPR materials is elements with a low diffusion coefficient. In this case, the Cr-rich and Cr-Fe phases were formed in all samples. The problem has not been overcome even if the milling process was performed for 40 hours, followed by heat treatment. Recorded results are promising, but more research in this area is badly needed.

**CRediT authorship contribution statement**

A. Olejarz: Conceptualization, Investigation, Formal analysis, Writing – Original draft. W. Y. Huo: Methodology, Writing – review & editing. M. Zieliński: Investigation, Formal analysis, Writing – review & editing. R. Diduszko: Investigation, Formal analysis. E. Wyszkowska: Investigation, Formal analysis. A. Kosińska: Investigation, Formal analysis. D. Kalita: Investigation, Formal analysis. I. Jóźwik: Formal analysis. M. Chmielewski: Investigation. F. Fang: Writing – review & editing. Ł. Kurpaska: Supervision, Writing – review & editing.

**Declaration of Competing Interest**

The authors declare that they have no known competing financial interests or personal relationships that could have appeared to influence the work reported in this paper.

**Acknowledgements**

   We acknowledge support from the European Union Horizon 2020 research and innovation program under NOMATEN Teaming grant (agreement no. 857470) and from the European Regional Development Fund via the Foundation for Polish Science International Research Agenda PLUS program grant No. MAB PLUS/2018/8.